\documentclass{optica-article}

\journal{opticajournal} % for journals or Optica Open

\articletype{Research Article}

\usepackage{bm}% bold math
\usepackage{here}
\usepackage{comment}

\usepackage{lineno}

\begin{document}

\title{Sub-photon accuracy noise reduction of single shot coherent diffraction pattern with atomic model trained autoencoder}

\author{Takuto Ishikawa,\authormark{1*} Yoko Takeo,\authormark{1,2} Kai Sakurai,\authormark{1} Kyota Yoshinaga,\authormark{1} Noboru Furuya,\authormark{1} Yuichi Inubushi,\authormark{2,3} Kensuke Tono,\authormark{2,3} Yasumasa Joti,\authormark{2,3} Makina Yabashi,\authormark{2,3} Takashi Kimura,\authormark{1} and Kazuyoshi Yoshimi\authormark{1}}

\address{\authormark{1}Institute for Solid State Physics, The University of Tokyo, Kashiwa-shi, Chiba 277-8581 Japan.\\ \authormark{2}Japan Synchrotron Radiation Research Institute, 1-1-1 Kouto, Sayo-cho, Sayo-gun, Hyogo 679-5198, Japan\\ \authormark{3}RIKEN, SPring-8 Center, 1-1-1 Kouto, Sayo-cho, Sayo-gun, Hyogo 679-5148, Japan}

\email{\authormark{*}takuto-ishikawa@issp.u-tokyo.ac.jp}

\begin{abstract*} 
Single-shot imaging with femtosecond X-ray lasers is a powerful measurement technique that can achieve both high spatial and temporal resolution. However, its accuracy has been severely limited by the difficulty of applying conventional noise-reduction processing. This study uses deep learning to validate noise reduction techniques, with autoencoders serving as the learning model. Focusing on the diffraction patterns of nanoparticles, we simulated a large dataset treating the nanoparticles as composed of many independent atoms. Three neural network architectures are investigated: neural network, convolutional neural network and U-net, with U-net showing superior performance in noise reduction and subphoton reproduction. We also extended our models to apply to diffraction patterns of particle shapes different from those in the simulated data. We then applied the U-net model to a coherent diffractive imaging study, wherein a nanoparticle in a microfluidic device is exposed to a single X-ray free-electron laser pulse. After noise reduction, the reconstructed nanoparticle image improved significantly even though the nanoparticle shape was different from the training data, highlighting the importance of transfer learning.
\end{abstract*}

%%%%%%%%%%%%%%%%%%%%%%%%%%  body  %%%%%%%%%%%%%%%%%%%%%%%%%%
\section{Introduction}
X-ray free electron lasers (XFEL)~\cite{Ishikawa2012-zg,Emma2010-cl}, a new generation of X-ray light sources, provide angstrom wavelengths, high peak brightness, and imaging on timescales shorter than 10 femtoseconds, a timescale faster than structural breakdown caused by the intense X-ray measurement~\cite{Neutze2000-hz,Chapman2006-mj,Seibert2011-bn,Kimura2014-rf}. This allows observation of nano-sized samples in their natural state, such as in aerosols~\cite{Seibert2011-bn,Schot_2015} or liquids~\cite{Matsumoto2022-ru,Suzuki2022-kd}, without radiation damage or sample immobilization while achieving higher spatial resolution than visible light microscopy. Its potential is expected to lead to breakthroughs in science and technology, such as the discovery of new material and biological structures. On the other hand, because this method uses a single-pulse measurement to obtain radiation-damage-free data, general noise reduction processing such as integration and background subtraction is not possible. This problem has been a barrier to achieve higher reliability and resolution.

Image noise reduction is a pervasive problem, which has been mitigated with various methods, such as filtering~\cite{4767641, 4271520, 10.1137/040616024} and sparse modeling methods~\cite{1614066, 4011956, JPSJ.89.012001}. Recently, deep learning (DL) has attracted attention.
Noise reduction using DL centers around compressing information to emphasize the inherent attributes of an image while stripping away unnecessary elements like noise. This approach is embodied in autoencoders (AE), which compresses and then reconstructs the input data. Various techniques have been developed based on AE, including the denoising AE~\cite{Vincent2008,Vincent2010}, which introduces noise to the input data and aims for image restoration, and the denoising convolutional neural network (DnCNN)~\cite{Zhang2017}, which incorporates CNN~\cite{GoodBengCour16,cnn_review2,CNN-review}, batch normalization~\cite{batch_normalization}, and residual learning~\cite{RL} in its NN architecture.

While DL offers advantages in image denoising tasks, it often faces bottlenecks due to its inherent need for large training data. However, this difficulty can be overcome by utilizing simulation data for supervised learning. Many XFEL experiments can be meticulously reproduced by simulation, creating an ideal situation for exploiting DL methods. This is because the phenomenon of the laser irradiating an object can be meticulously reproduced by simulation. In fact, 
applications of DL methods to coherent diffraction patterns with XFEL have been reported for simple noise reduction~\cite{Lee_2021}, phase recovery~\cite{phase_recovery_2d_2018,phase_recovery_2d_2020,phase_factor_dl}, and super-resolution~\cite{super_resolution_by_dl}. 

In this report, we propose noise reduction of XFEL-Coherent Diffractive Imaging (CDI) single-shot data, which is difficult to process in the background, using DL methods.
To generate a large amount of simulation data for arbitrarily-shaped objects, the diffraction pattern data is generated by atomically-graining the sample particles and calculating the structure factor using parallel computing. Using this dataset, we compare the efficiency of the denoising process among three networks (NN, CNN, and U-net~\cite{Unet}). 

The structure of this paper is as follows. In Sec.~2, we provide a detailed description of how the simulated data are created using experimental data. We also describe our comprehensive approach to preparing the necessary training dataset for DL. In addition, this section delves into the specifics of the architectures for NN, CNN, and U-net, detailing the conditions under which the actual computations are performed. In Sec.~3, we describe the training dataset, the results and evaluation of training with NN, CNN, and U-net, and the results of applying noise reduction to the artificial data. This section is complemented by the results of transfer learning applications to different shapes and structures and a presentation and evaluation of noise reduction when applied to the experimental data. Finally, we summarize the key findings of this study in Sec.~4.
\begin{figure}[H]
\centering\includegraphics[clip,width=13.0cm]{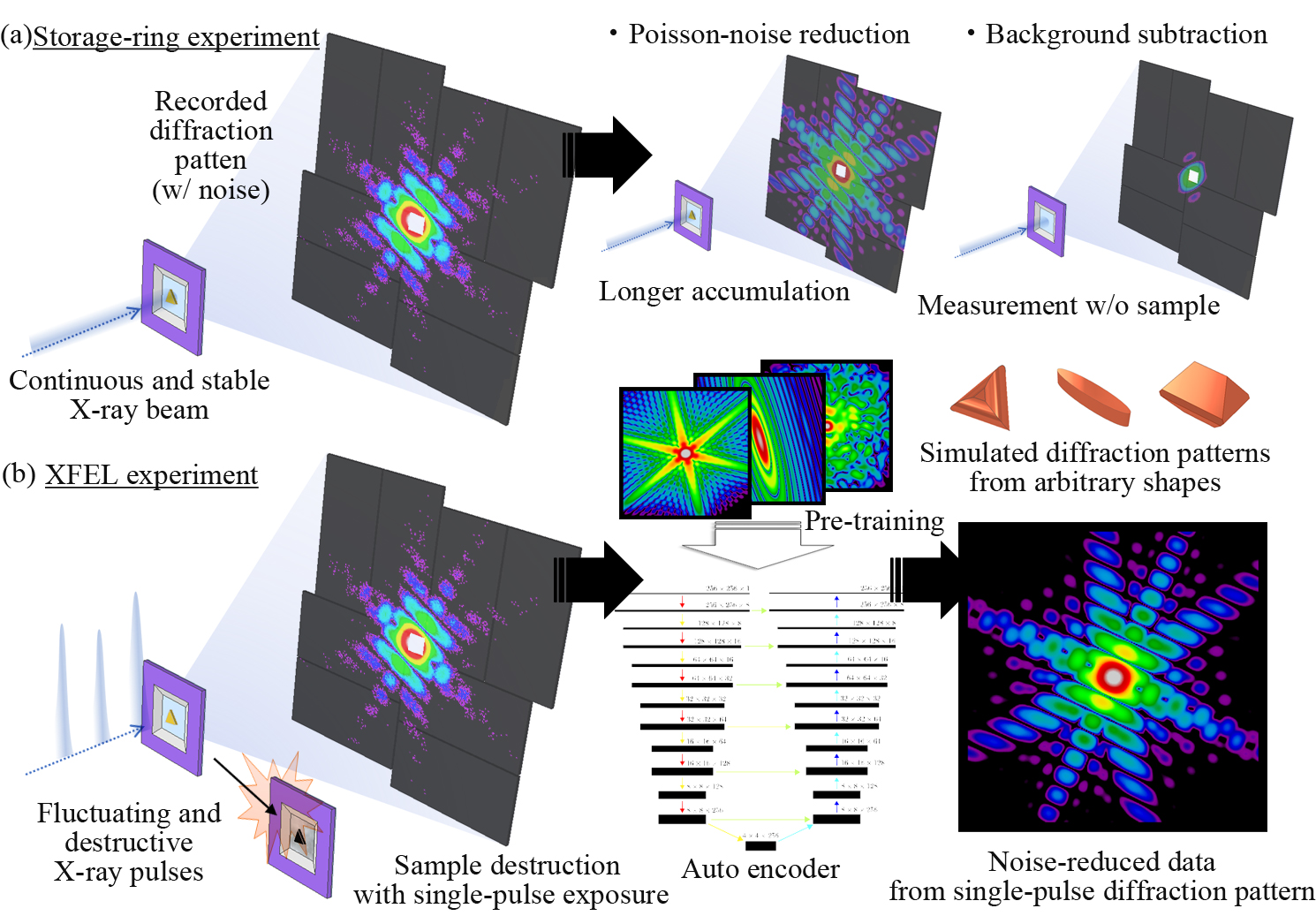}
\caption{Schematic of noise reduction processes in coherent diffraction imaging.
(a) In a typical synchrotron radiation experiment, Poisson noise is reduced by increasing the exposure time,
and background noise is subtracted by measuring the state without a sample.
(b) In the case of XFEL experiments, noise reduction is difficult because the beam profile fluctuation is large, and the sample is destroyed after a single-shot exposure.
Therefore, noise reduction is performed with an autoencoder pre-trained with noiseless diffraction patterns derived from simulated data.}
\label{fig:use_image}
\end{figure}

\section{\label{sec:method}Methods}
In conventional storage ring experiments, a nearly continuous and stable incident X-ray beam is applied to the sample as shown in FIG.~\ref{fig:use_image} (a). Therefore, noise in the coherent diffraction patterns can be reduced by long exposures or by removing background signals by measuring without the sample.
In contrast, XFEL measurements have large shot-to-shot variations due to the X-ray generation process starting from noise in a process termed self-amplified spontaneous emission. This strong variation makes it challenging to apply conventional noise removal methods. In addition, for single-pulse measurements using the diffraction-before-destruction scheme with a focused XFEL beam, the sample and its holder are destroyed in a highly intense single-pulse exposure, making statistical processing and background measurement impossible.
Therefore, for such single-pulse XFEL measurements, we have developed an AE-based denoising method that learns accurate simulated diffraction pattern data from an atomic model of nanostructures as shown in FIG.~\ref{fig:use_image} (b).
To effectively apply DL techniques to image denoising, a systematic approach to constructing training datasets and model architectures is essential. 
In the following, we explain the method to generate the simulation dataset and DL architecture used in this study.

\subsection{Generation of simulated dataset}
As shown in FIG. 2(a), the first step in generating a training dataset is to generate noise-free simulated images using the coherent diffraction intensity distribution function. These noise-free images are then subjected to various transformations, such as zooming, rotation, and shearing to produce a comprehensive set of correct data. Further details of these transformations can be found in Sec.~3.1.
Once the training dataset is generated, we construct the DL model. As shown in FIG. 2(b), the architecture can be based on a NN, a CNN or a U-net. To be consistent with the experimental setup, we add Poisson noise, discretize the image for photon counting, and apply a window considering the detector. If the window is left in place, the learning accuracy will be degraded, so we interpolate the window. This data is put into the input, and the output is trained to recover the noiseless ground truth.
Finally, after finishing training, the model becomes adept at denoising. When provided with experimental data containing noise, as shown in FIG. 2(c), the model effectively outputs a denoised version of the input image.

\begin{figure}[H]
\centering\includegraphics[clip,width=10.0cm]{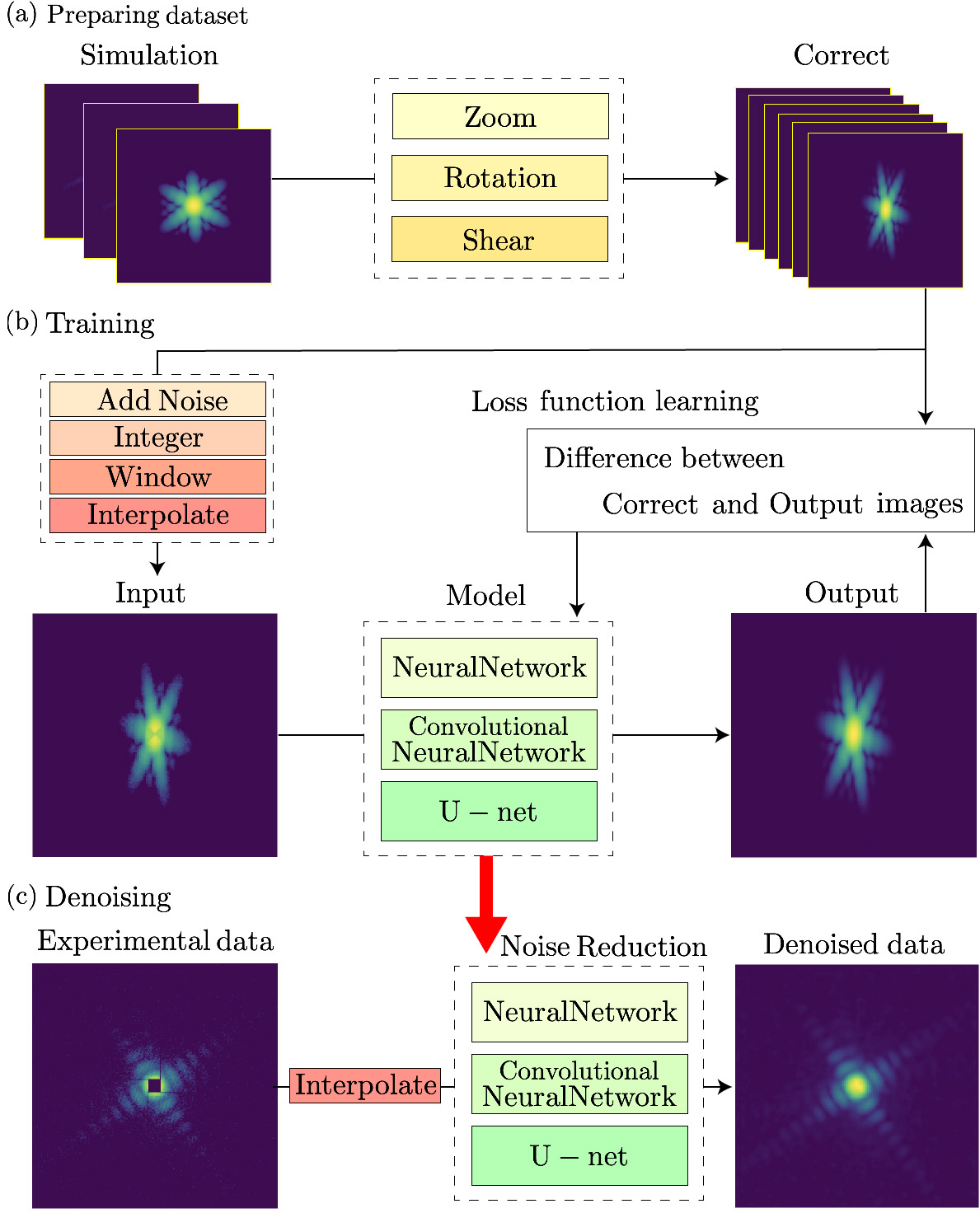}
\caption{\label{fig:ae} Schematics for constructing training models to denoise images. (a) We first obtain the images without noise using the coherent diffraction intensity distribution function to generate the training dataset. The data is corrected by applying affine transformations. (b) We construct a training model using a NN, CNN, or U-net that input data created from the ground truth by adding noise, discretizing pixel values, windowing, and interpolating. The output data is trained to return the ground truth without these steps. (c) Using the training model created in (b), we recover a denoised image as output data when the experimental data is inputted and interpolated. Details of transformation processes and image processing are described in Sec.~3.1.
}
\end{figure}
XFEL-CDI experiments can be modeled as coherent far-field diffraction, resulting in a diffraction pattern that is close to the two-dimensional Fourier transform of the illuminated object.
The coherent diffraction intensity distribution function $I({\bm{K}})$ on a two-dimensional detector can be calculated by using the electron density distribution function of the material $\rho ({\bm{r}})$.
\begin{align}
    I({\bm{K}}) =  I_{0} \frac{P r_{\rm e}^{2}}{l^{2}}
    \left|\int \rho ({\bm{r}}) e^{i {\bm{K}} \cdot {\bm{r}} } d{\bm{r}} 
    \right|^{2},
    \label{eq:Ik_1}
\end{align}
where $I_{0}$ is the intensity of incident X-ray, $\bm{K}$ is the wave vector, $\bm{r}$ is the position,  $r_{\rm e}$ is the classical electron radius, 
$P$ is the polarizing factor, and $l$ is the distance between the sample and the detector. 
In preparing the simulation data, we divide the target material into small spheres as individual atoms to represent its shape for simplicity. In this case, $\rho(\bm{r})$ is defined as $\sum_{i} f_i \rho_0(\bm{r}-\bm{R}_i)$. Then, Eq. (\ref{eq:Ik_1}) can be rewritten as
\begin{align}
    I({\bm{K}}) &= I_{0} \frac{P r_{\rm e}^{2}}{l^{2}}
    \left|\sum_{i} f_i e^{i{\bm K}\cdot{\bm R}_i}\right|^{2} \left|F(K) \right|^{2},
     \label{eq:Ik_2}
\end{align}
where $F(K)\equiv \int \rho_0 (\bm{r}) e^{i {\bm{K}} \cdot \bm{r}} d{\bm{r}}$ with $K=|{\bm K}|$ is a structure factor for a sphere. In our calculation, we set $\rho_0(\bm{r})$ as $\rho_0$ for $|\bm{r}|<r_0$ and $0$ for $|\bm{r}|>r_0$. Using this relation, $F(K)$ is given as
\begin{align}
F(K) = \frac{4 \pi \rho_0}{K^3} \{ -K r_0 \cos(K r_0)+\sin(K r_0) \}.
\end{align}

In this study, we first generate simulation data of coherent diffraction patterns arising from nanoparticles based on the above formula.
To generate diffraction patterns for different rotations of the object, we apply affine transformations to the final image.
Details of this procedure are described later (Sec. 3.1).

\subsection{Deep-Learning architecture}
\begin{figure}[H]
\centering\includegraphics[clip,width=10cm]{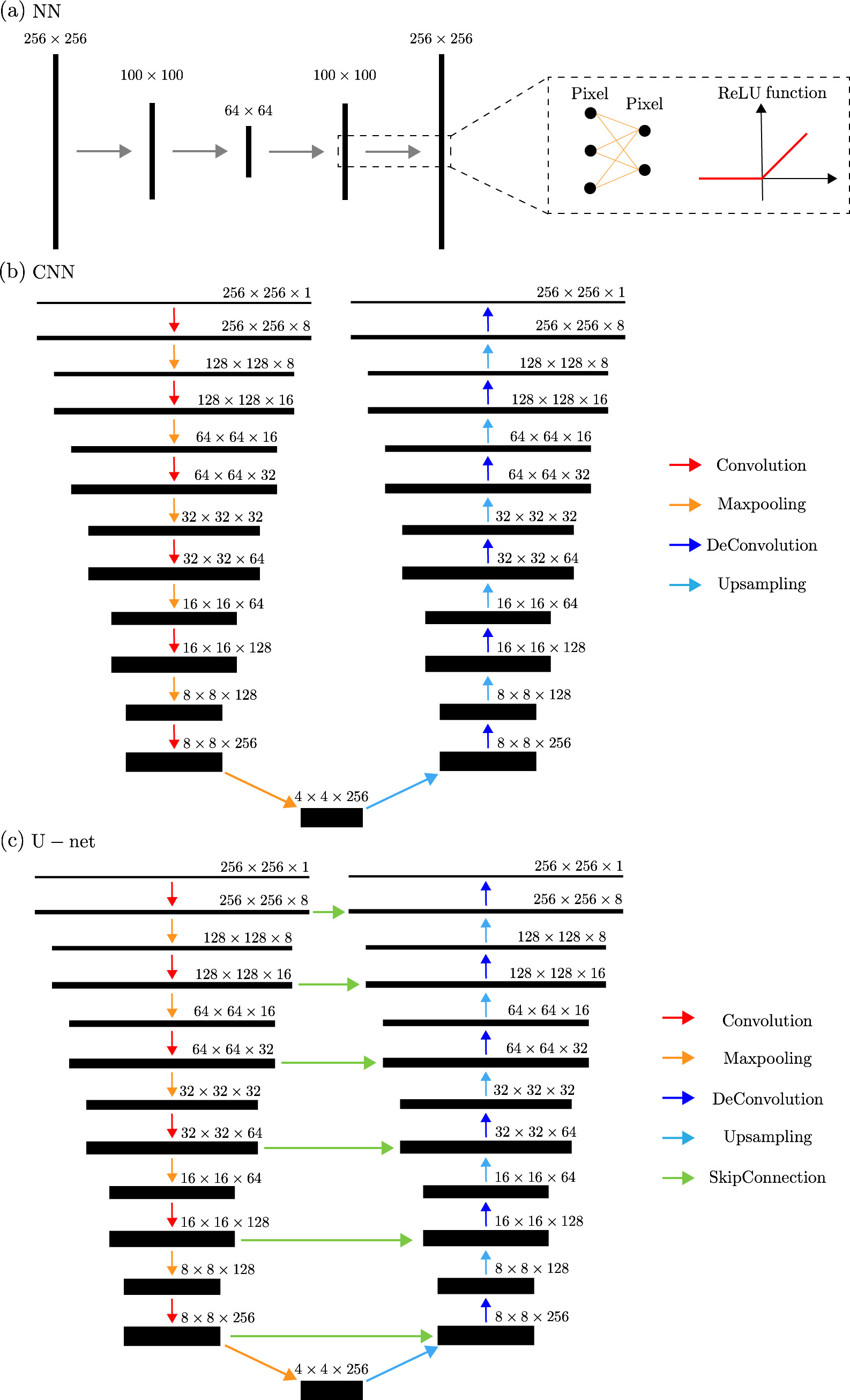}
\caption{DL architecture for (a) NN, (b) CNN, and (c) U-net in this study. 
(a) Each layer is fully connected, and each uses the ReLU function as an activation function. (b) The red arrow indicates the convolution layer and it plays a role in extracting image features. The orange arrow indicates the pooling layer and its compressed information. Blue and sky blue arrows indicate de-convolution and up-sampling. They are the inversion of convolution and pooling. The activation function is the ReLU function. (c) This model is based on CNN, but it contains a skip connection (green line) in the same dimension. Skip connection is fully connected, and it avoids the lack of details coming from compressed information.
}
\label{fig:dl}
\end{figure}

-- {\it Neural Network (NN)}

NN is known as the core architecture for the DL method. To reduce noise from the image data, we construct the AE shown in FIG.~\ref{fig:dl}(a). In this study, the dimension of the input data $y$ is $256 \times 256$, equal to the total number of pixels for the training data image. To extract the features of the image, the dimension of layers is gradually reduced from $256\times 256$ to $100\times 100$, and finally to $64 \times 64$. Then, to decode the image, the dimension of layers is gradually increased from $64\times 64$ to $100\times 100$, and finally to $256 \times 256$, which is the decoded data ${\hat y}$. In our calculation, we perform mini-batch learning with a batch size of 20. The total parameter number of this network is $1,392,640,000$. 
\\
-- { \it Convolutional Neural Network (CNN) }

CNN is known to be better suited than NN for capturing object shapes through filters (kernels) in images. Figure \ref{fig:dl}(b) shows the CNN AE considered in this study. We use $8, 16, 32, 64$, and $128$ kernels that operate on $3 \times 3$ pixels of a 2D image for each layer to capture the shape of images. These kernels have nine variables, so they are learned as parameters during training. By increasing the number of kernels, various image features can be extracted. The maxpooling layer compresses $L \times L$ images into $L/2 \times L/2$ images by outputting the maximum value of $2 \times 2$ pixels in 1 pixel. The upsampling layers add dimensions to the image by padding white space to the image and then using the kernel.
Input figure dimension is $256 \times 256 \times 1$ and compressed to $4 \times 4 \times 256 = 4096$ dimension using convolutional and maxpooling layers. Then, to decode the image, the dimension of layers gradually increases by deconvolution and upsampling, and finally, for $256 \times 256 \times 1$ which is the decoded data ˆy.
We perform mini-batch learning with a batch size of 20. The total Parameter number of this network is $3,152,641$.
\\
-- { \it  U-net}

Though CNN is good at capturing object features, it tends to average out its microstructure. U-net is one method to overcome this weakness. The difference from CNN is the existence of contracting paths that connect between the encoding layer and decoding layer. Through the contracting path, the detailed location information can be preserved. Figure \ref{fig:dl}(c) shows the U-net considered in this study. In the encoding layers, the architecture is the same as CNN and input data whose dimension is  $256 \times 256 \times 1$ compressed to $4 \times 4 \times 256 = 4096$ dimension using convolutional and maxpooling layers. But in decoding layers, as the input layers of the deconvolution, the image of the previous layer and the image of the same dimension obtained by the encoding are used. The path used to decode the encoded data is called the contracting path and is shown in the green arrow in Fig.\ref{fig:dl}(c). We perform mini-batch learning with a batch size of 20. The total Parameter number of this network is 4,731,673.

To optimize parameters included in the NN, CNN, and U-net models, we use adaptive moment estimation (ADAM)\cite{adam} and a rectified linear unit (ReLU) function~\cite{ReLU} as the activate function. As a loss function, we used the mean squared error, MSE$\equiv \frac{1}{n} \sum_{i=1}^{n} (\hat{y}-y)^2$ where $\hat{y}$ is decoded data, and $y$ is correct data, $n$ is a total number of pixels in the image. In our calculation, the following GPUs and software packages are used: Two GPU NVIDIA Corporation GA100 [A100 SXM4 40GB], python version 3.10.4, tensorflow-gpu 2.9.1, cudatoolkit 11.3.1, cudnn 7.2.1, numpy 1.23.0, scipy 1.7.3, and keras 2.9.0.

\section{\label{sec:result}Results}
In this section, we first introduce the details of the training datasets. Then, we describe the details of the learning process of the denoising model using the NN, CNN, and U-net and evaluate their performances. Next, we apply the trained denoising model to shapes different from those used in training. The effects of transfer learning on learned models will also be evaluated. Finally, we discuss the efficiency of the denoising for experimental data using the trained model, and also discuss reconstruction of the denoised images.

\subsection{Training dataset}
\begin{figure}[H]
\centering\includegraphics[clip,width=13.0cm]{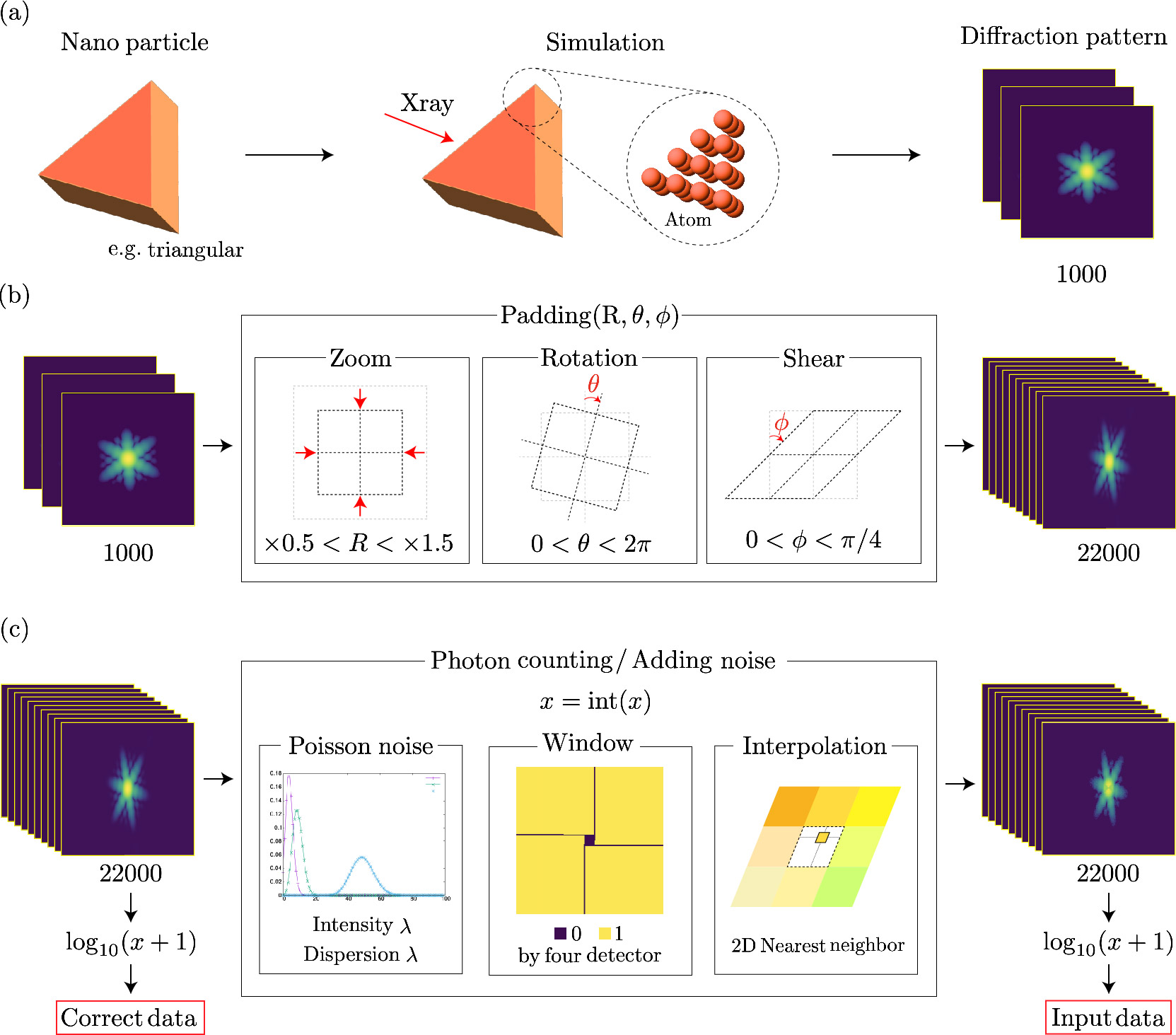}
\caption{
Schematic of the training data generation process. (a) Generation of diffraction patterns from nanoparticles coarse-grained into atoms. (b) Image manipulation for changes in particle shape was performed to increase the number of diffraction patterns for training, such as zoom, rotation, and shear transformation.  (c) Rounding simulation to count single photons, adding noise, and windowing filters to approximate the experimental setup. Interpolation over the detector gaps to avoid edge artifacts in the learning model.
}
\label{fig:pre_training}
\end{figure}

As shown in FIG.~\ref{fig:pre_training} (a),  for creating the training data, we first calculated $I(K)$ according to Eq.~(\ref{eq:Ik_2}) for a material with a triangular shape, where $P$, $l$, and $I_0$ are set as $1.0$, $0.32$ m, and $6\times10^{11}$ photons/(100 nm)$^{2}$, respectively. These parameters are based on the usage of SACLA's CDI system using 100-nm focusing system\cite{Yumoto_2022}. In this study, we simulated 1,000 different orientations of the nanoparticle, producing images of 256x256 pixels.
With DL, the amount of training data is crucial for the learning process, and the amount of training data is often padded by processing the original images.
By considering the detector geometry, we scaled the images by randomly varying the magnification $R$ between $0.5$ and $1.5$. To rotate the images,  the rotation angle $\theta$ was randomly moved between $0$ and $2\pi$, and the image was padded. In addition, we applied shear transformations by randomly varying the angle $\phi$ between $0$ and $\pi/4$ and padded the image. In FIG.~\ref{fig:pre_training} (b), we show the schematic of the training dataset padding procedure. The images were padded from 1,000 to 22,000 by these processes. We used 20,000 images for the training dataset, and 2,000 images for the validation dataset.

In order to match the simulation data generation conditions to the experimental setup, the simulation data were generated as photon number and a window was applied to the data as shown in FIG.~\ref{fig:pre_training} (c).
This is because that the detector consists of eight panels arranged to avoid measuring the central intensity of the X-rays and realize low noise levels to achieve single photon detection.
Then, Poisson noise is added to account for the photon counting statistics.
The Poisson distribution is widely used to represent the number of events randomly occurring in a fixed period or space. The following equation defines this distribution:
\begin{align}
P(X=k) = \frac{\lambda^{k} e^{-\lambda}}{k !}.
\label{eq:poisson}
\end{align}
Here, the parameter $\lambda$ represents the average number of events per interval. A notable characteristic of the Poisson distribution is that its mean and variance are equal to $\lambda$. This property applies particularly when events arise independently at a constant average rate. We set $\lambda$ as photon intensity.

Reconstructing the original object requires accurate information about the interference fringes of the diffraction pattern $I({\bm K})$.
To reduce the influence of the detector edges on learning, diffraction patterns are interpolated with nearest neighbor interpolation.
For training images, logarithmic scale intensities were used to handle weak-intensity data to obtain information on the interference fringes. 

\subsection{Training process of denoising models}

\begin{figure}[H]
\centering\includegraphics[clip,width=13.0cm]{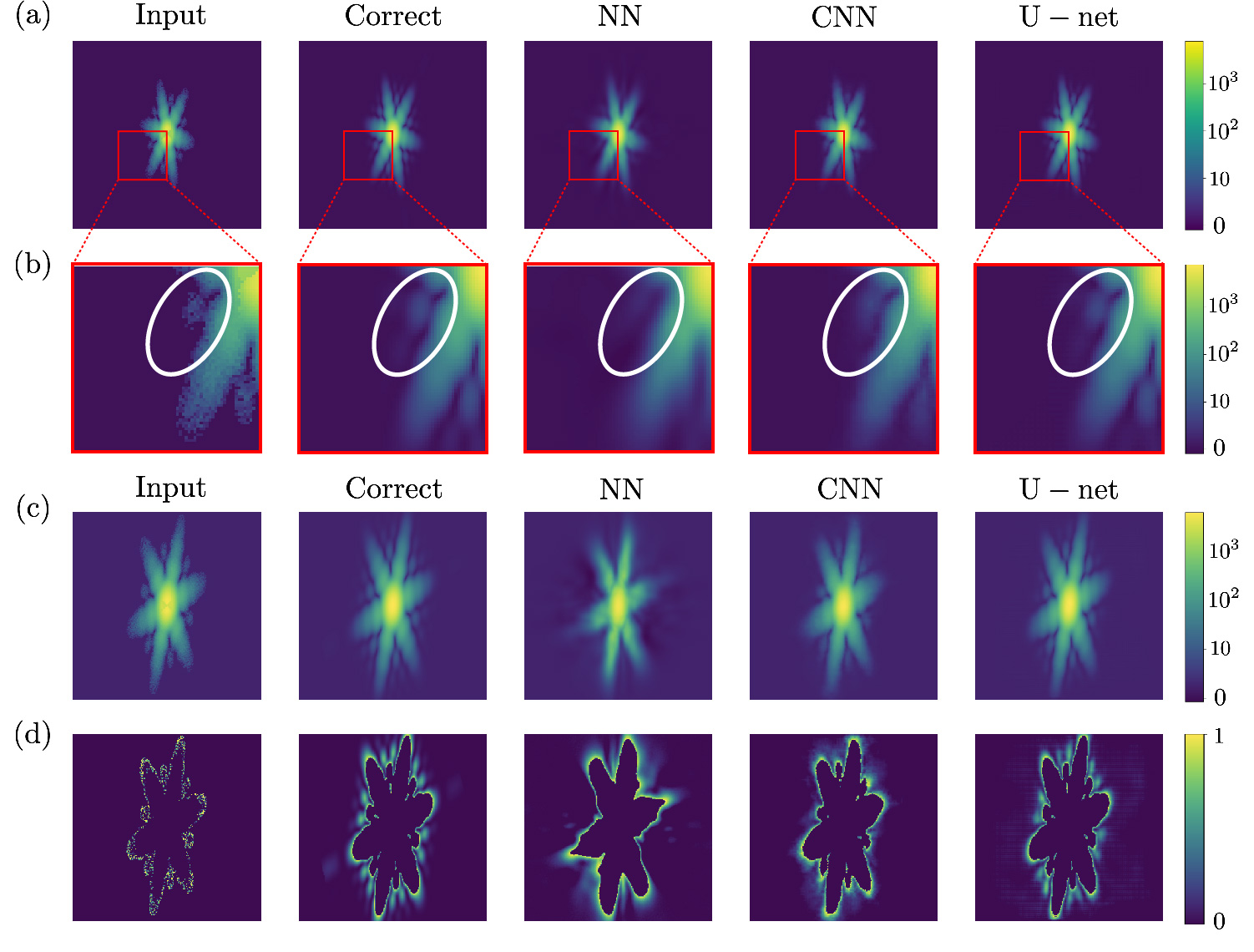}
\caption{
The denoising results obtained by each networks. 
In each row, images are arranged from left to right, showing Input, Correct, NN, CNN, and U-net. 
(a) The decoded patterns. (b) An enlarged view of row (a) highlighting one detailed interference fringe circled in white. (c) The decoded patterns for seeing the reconstructing effect of subphotons. (d) An enlarged images with photon counts below one. Regions with photon counts exceeding one and areas with zero photons are distinctly marked in blue.
}
\label{fig:train_1} 
\end{figure}

Figure ~\ref{fig:train_1} show the training results using simulation data. The first column (Input) corresponds to the generated simulation data with noise, the second column (Correct) shows simulation data before application of noise, and the third (NN), fourth (CNN), and fifth (U-net) columns are the decoded data using the learned model. The color scale is in photon number

All networks performed well where the intensity is strong, but significant differences were observed in features with subphoton intensity including weak interference fringes.
To see the effects of weak interference fringes clearly, an enlarged view focusing on low-intensity areas is provided in FIG.~\ref{fig:train_1} (b).
We can see that the NN-learned model is able to reproduce shapes, but misses the details associated with the interference fringes. In contrast, both the CNN and U-net models are able to reproduce shapes together with the interference fringes. In particular, the U-net model excels at recovering detailed interference fringes and shows superior performance in noise removal.
Figures.~\ref{fig:train_1} (d) show the results restricted to less than one photon. Compared to NN and CNN, the image obtained by U-net is closer to the correct one. Since the input data is discretized to count individual photons, it contains no intensity information in regions less intense than a single photon. Nevertheless, the U-net model can recover this subphoton information.

\begin{figure}[H]
\centering\includegraphics[clip,width=11.0cm]{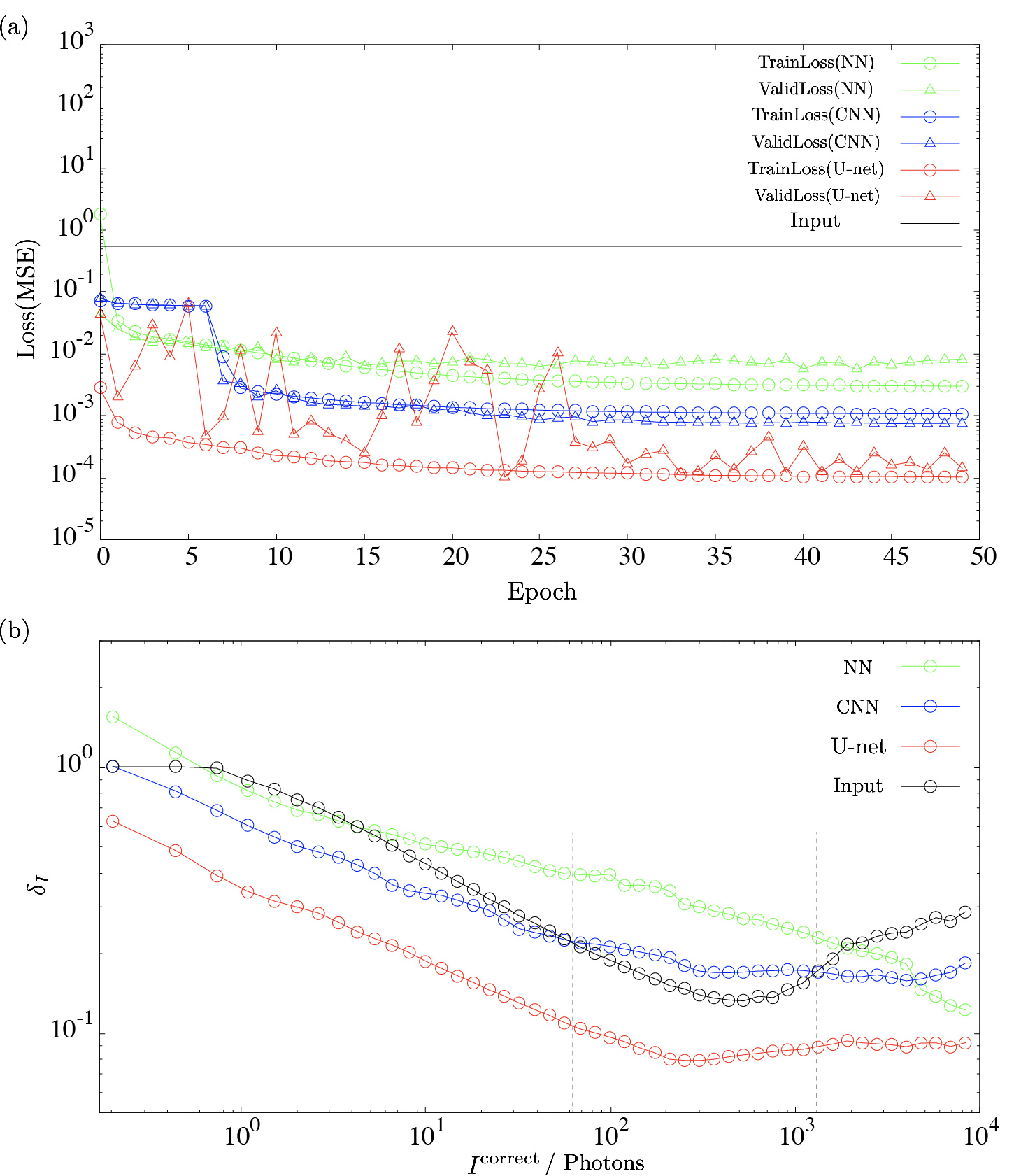}
\caption{\label{fig:train_2} (a) Epoch dependence of the loss function calculated by MSE. The green, blue, and red colors correspond to the data by NN, CNN, and U-net, respectively. The circle and triangular symbols correspond to the loss function obtained by the training data and validation data, respectively. (b) MSE for each intensity.
The calculation was performed for intensities $10^{0.1n}$, where n is an integer from 1 to 40.
The green, blue, red, and black colors correspond to the data by NN, CNN, U-net, and Input, respectively. The dot lines are guides to the eye for seeing the region where the CNN data is worse than the input data.}
\end{figure}

Figure~\ref{fig:train_2} (a) shows the epoch-dependence of the MSE loss function for training data (circle symbol) and validation data (triangular symbol), respectively. The green, blue, and red symbols correspond to the numerical results obtained by NN, CNN, and U-net, respectively. 
The MSE calculated using the input data without training is plotted with a solid black line. The values of MSE for all methods gradually decrease with increasing epochs, and the values converge approximately when the number of epochs exceeds around $30$. 
The training MSE and validation MSE for each network converge to approximately the same value, indicating that each network is not  over-trained. The final converged values are $3.05 \times 10^{-3} $ for NN, $1.09 \times 10^{-3}$ for CNN, and $1.04 \times 10^{-4}$ for U-net. This result indicates that the U-net shows the best noise reduction performance in this case.
To understand how the pixel intensity affects the accuracy of the denoising, Fig. 6 (b) shows a histogram of the intensity-normalized MSE over pixel intensity. We define the intensity-normalized MSE to be
\begin{align}
    {\delta}_{I}(a_1;a_2) = \frac{1}{N} \sum_{i \in (a_1 <  I^{\rm{correct}}_i  <a_2) } (1-I^{\rm{decoded}}_i/I^{\rm{correct}}_i)^2,
\end{align}
where $(a_1;a_2)$ means the intensity bin within the interval from $a_1$ to $a_2$, $N$ is the total number of pixels in the intensity bin, $i$ is the pixel index, and $I_i^{\rm correct}$ and $I_i^{\rm decoded}$ are pixel intensities from the correct (decoded) data.

The MSE for input data (${\rm MSE}_{\rm Noise}$) approaches 1 below $I^{\rm{correct}} < 1$ as the intensity falls below the photon discretization. As the intensity increases, ${\rm MSE}_{\rm Noise}$ decreases and turns to increase from  about $4 \times 10^2$ photons. The increase in ${\rm MSE}_{\rm Noise}$ is due to windowing and interpolation errors. 
The MSE for NN data (${\rm MSE}_{\rm NN}$) tends to decrease monotonically with increasing intensity. However, in the intensity range from $4\times10^0$ photons to $2\times10^3$ photons, ${\rm MSE}_{\rm NN}$ has higher value than ${\rm MSE}_{\rm Noise}$, which corresponds to the disappearance off the interference fringes in the NN model reconstruction in FIG. \ref{fig:train_1}.
The MSE for CNN data (${\rm MSE}_{\rm CNN}$) tends to decrease with increasing intensity and saturates around $3\times 10\time2$ photons.
Overall, the results show an improvement compared to the ${\rm MSE}_{\rm NN}$, except in areas of high intensity.
This improvement leads to the recovery of the interference fringes as seen from FIG. \ref{fig:train_1}.
Like as ${\rm MSE}_{\rm CNN}$, the MSE for U-net data (${\rm MSE}_{\rm Unet}$) tends to decrease with increasing intensity and saturates around $3\times 10\time2$ photons. 
However, ${\rm MSE}_{\rm Unet}$ has the lowest MSE than other methods and ${\rm MSE}_{\rm Unet}$ is lower than ${\rm MSE}_{\rm input}$ for all intensities.
This indicates that interference fringe features are efficiently learned by using the skip connection method. In particular, it can be seen that even in the sub-photon intensity region, ${\rm MSE}_{\rm Unet}$ is the lowest. These improvements led to the high-quality reconstruction by U-net, as shown in FIG. \ref{fig:train_1}.

Finally, we evaluate the efficiency of denoising when the model learned with U-net for the triangle shape is applied to different shapes such as ellipses and random-shaped particles, and transfer learning~\cite{transfer_learning_survey}, which is a method of increasing the speed and accuracy of learning by setting the parameters of one learning model as initial values of another.

\begin{figure}[H]
\centering\includegraphics[clip,width=13.0cm]{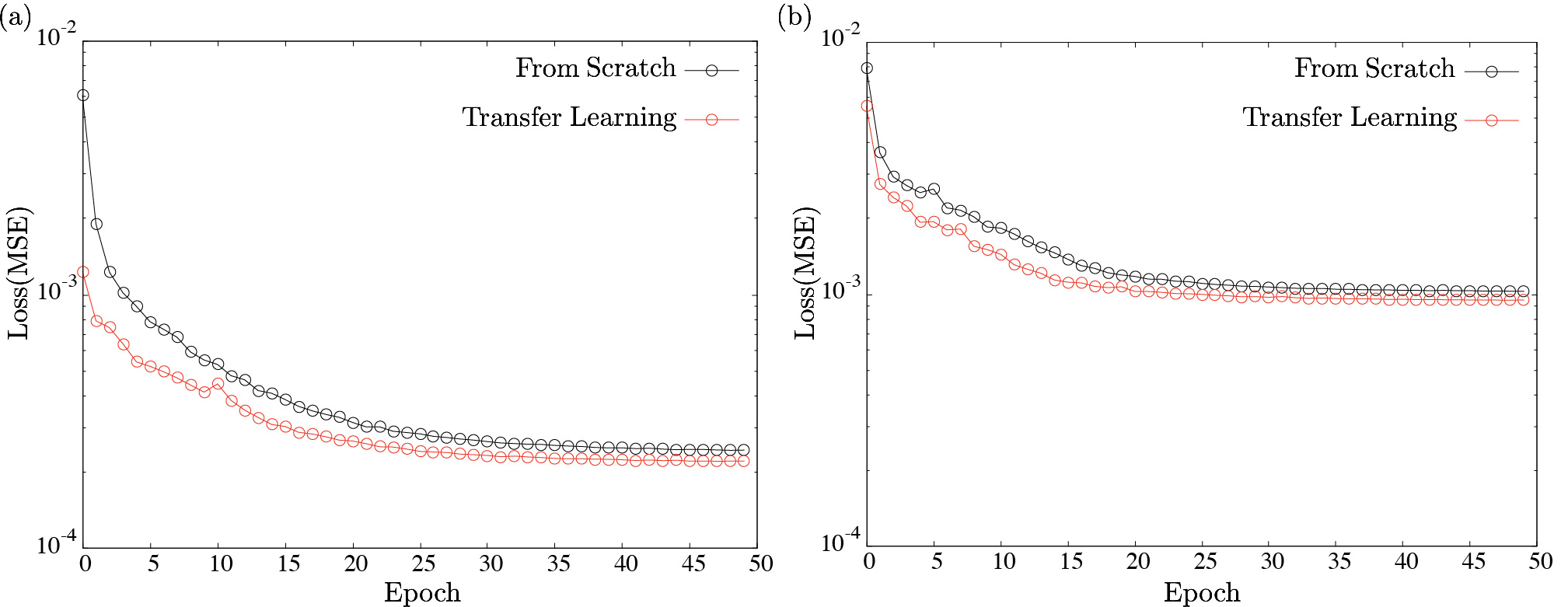}
\caption{\label{fig:transfer}
The epoch dependence of loss functions with U-net for (a) ellipse particles and (b) random-shaped particles.
The results from scratch, trained from random initial values, are shown as black symbols. The results of transfer learning with the learned triangle parameters as initial values are shown as red symbols. 
}

\end{figure}

Figures \ref{fig:transfer} (a) and (b) show the epoch dependence of the MSE between the correct and decoded data for the ellipse particles and random-shaped particles, respectively. The black circles show the training results from scratch, and the red circles show the results of a fine-tuned model trained in a triangular shape as a starting point.
For both shapes, when compared to models trained from scratch, the performance accuracy was found to be almost equivalent. However, a distinct advantage of utilizing transfer learning is observed in convergence speed, enabling models to reach optimal performance faster.

\begin{figure}[H]
\centering\includegraphics[clip,width=13.0cm]{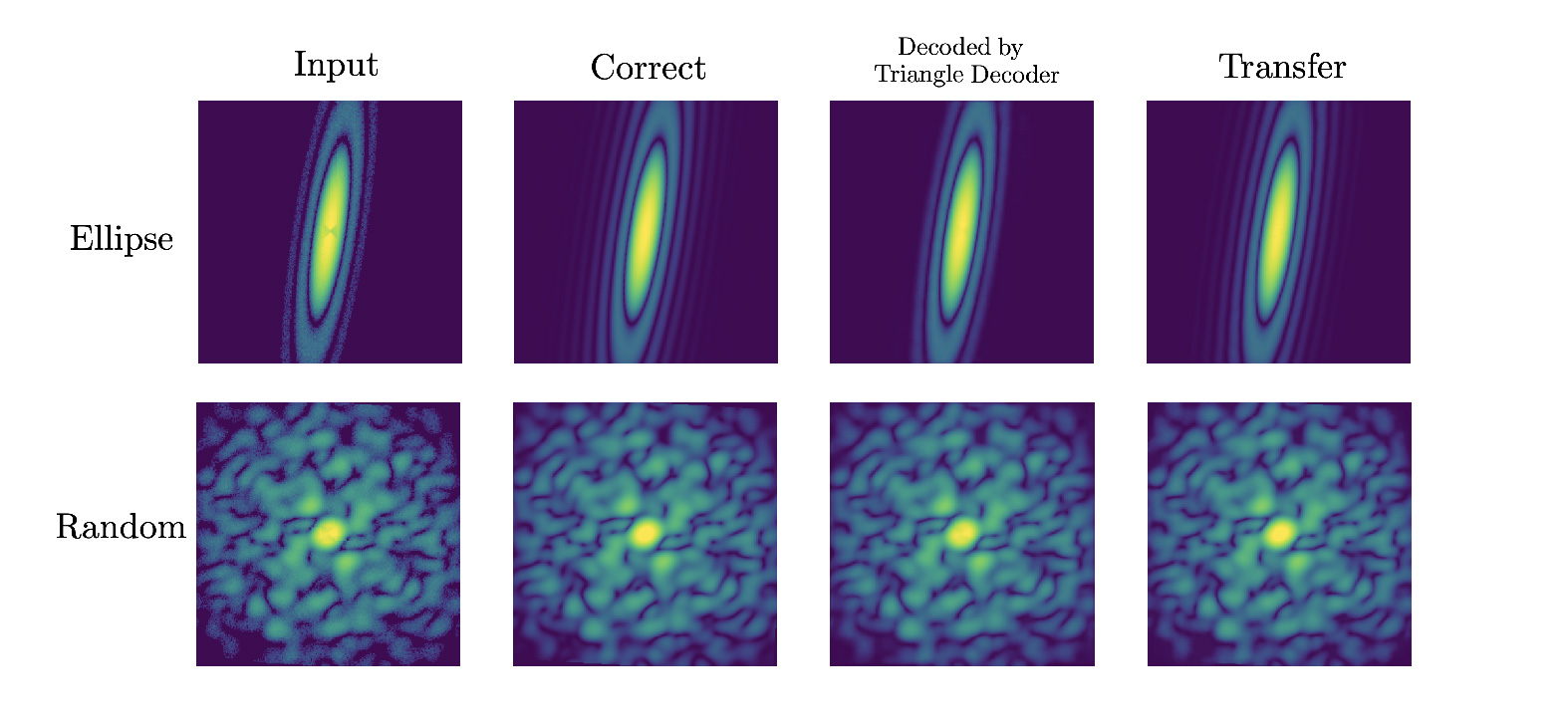}
\caption{\label{fig:othershape}
The denoising images were obtained using pre-trained and fine-tuned models. The top row is for the elliptical shape, and the second is for the random shape. For both rows, from left to right, the images represent the input image (with noise added), the correct image (without noise), the decoded image using a pre-trained model for triangular particle shapes, and the decoded image using a fine-tuned model, respectively.
}
\end{figure}

Figures \ref{fig:othershape} show the decoded images when using the pre-trained model and the fine-tuned model. The pre-trained model manages to reconstruct the denoised image for elliptical shapes, but its accuracy is limited to the third interference fringe, resembling the noise-infused image. In contrast, the transfer-learned model adeptly recreates interference fringes up to the fifth, achieving a near-perfect alignment with the reference image. 
For random shapes, both methods exhibited comparable performance, indicating minimal differentiation. This highlights that while transfer learning provides acceleration benefits in training, its superior performance might be more pronounced for specific shapes like ellipses but less for messy shapes.

\subsection{\label{sec:application} Reconstruction of denoised images}

\subsubsection{\label{sec:reconstruction}Application to simulation data}

In order to reconstruct the denoised X-ray data into a real space picture, a phase recovery calculation is needed. We used the relaxed averaged alternating reflections algorithm \cite{relaxed} and the shrink-wrap algorithm \cite{shrink-warp} to retrieve sample images from the coherent diffraction patterns. We performed 200 runs of real-space image reconstructions from different initial random seeds for each diffraction pattern. When reconstructing the experimental real-space image, a masking process is used to interpolate the blank areas caused by the detector geometry.

\begin{figure}[H]
\centering\includegraphics[clip,width=13.0cm]{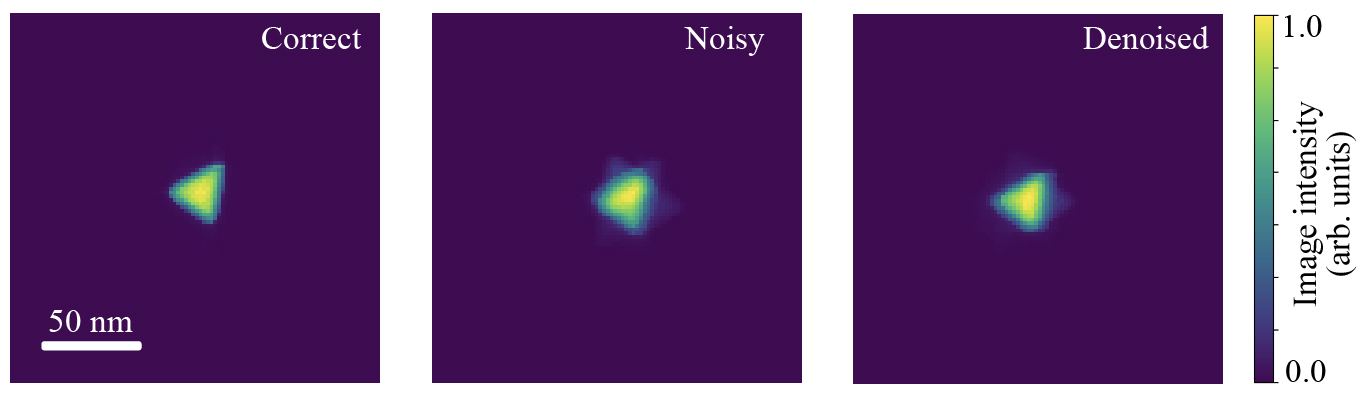}
\caption{\label{fig:sim_noise_reduction} 
Results of reconstructing simulation data for the triangular particle. Correct, Noisy, and Denoised data are the data generated by the simulation data corresponding to Fig.\ref{fig:pre_training} (a), the simulation data with the noise corresponding to Fig.\ref{fig:pre_training} (b), and the data with noise reduction by the U-net model, respectively.
}
\end{figure}

Figure \ref{fig:sim_noise_reduction} shows the averages of all the reconstruction results from simulated diffraction patterns under different conditions.
To evaluate the accuracy of the reconstruction data, 
the following three types of data are used: the simulation data itself, the simulation data with noise, and the denoised data using the U-net model.
The denoised data shows clear shapes and contours, which are consistent with the reconstruction results of the simulated data, but the noisy data is still obscure compared to the denoised data.

\subsubsection{\label{sec:experiment} Application to experimental data}
XFEL-CDI can observe a radiation-damage-free structure of a sample with a femtosecond single pulse exposure from an XFEL. AgCl nanoparticles are representative photosensitive materials, which are easily deformed during visible light or electron beam measurement~\cite{Formo-2012}. Here, we applied noise reduction processing using a learned noise eliminator to observe the natural structure of AgCl nanoparticles in a microfluidic device with a single-pulse XFEL. In the following, we first describe the experimental setup, and then the performance of our method is verified by denoising the experimental images.

The XFEL-CDI experiments were performed at the hard X-ray beamline BL2 of the SPring-8 Angstrom Compact Free-electron Laser (SACLA~\cite{Ishikawa2012-zg}). We used the high-resolution coherent diffraction pattern system MAXIC-S\cite{Yumoto_2022}, permanently installed at the beamline. The experimental photon energy was 4.0 keV. We used a multi-port charge-coupled device (MPCCD~\cite{Kameshima_2014}) with single photon counting sensitivity for two-dimensional detection of coherent diffraction patterns. AgCl nanoparticles were used as a demonstration sample. The typical shape of the nanoparticles is shown in Figure 10(a). The measurements were performed in the dispersed state in solution using a microfluidic solution sample holder~\cite{Matsumoto2022-ru}.

\begin{figure}[H]
\centering\includegraphics[clip,width=13.0cm]{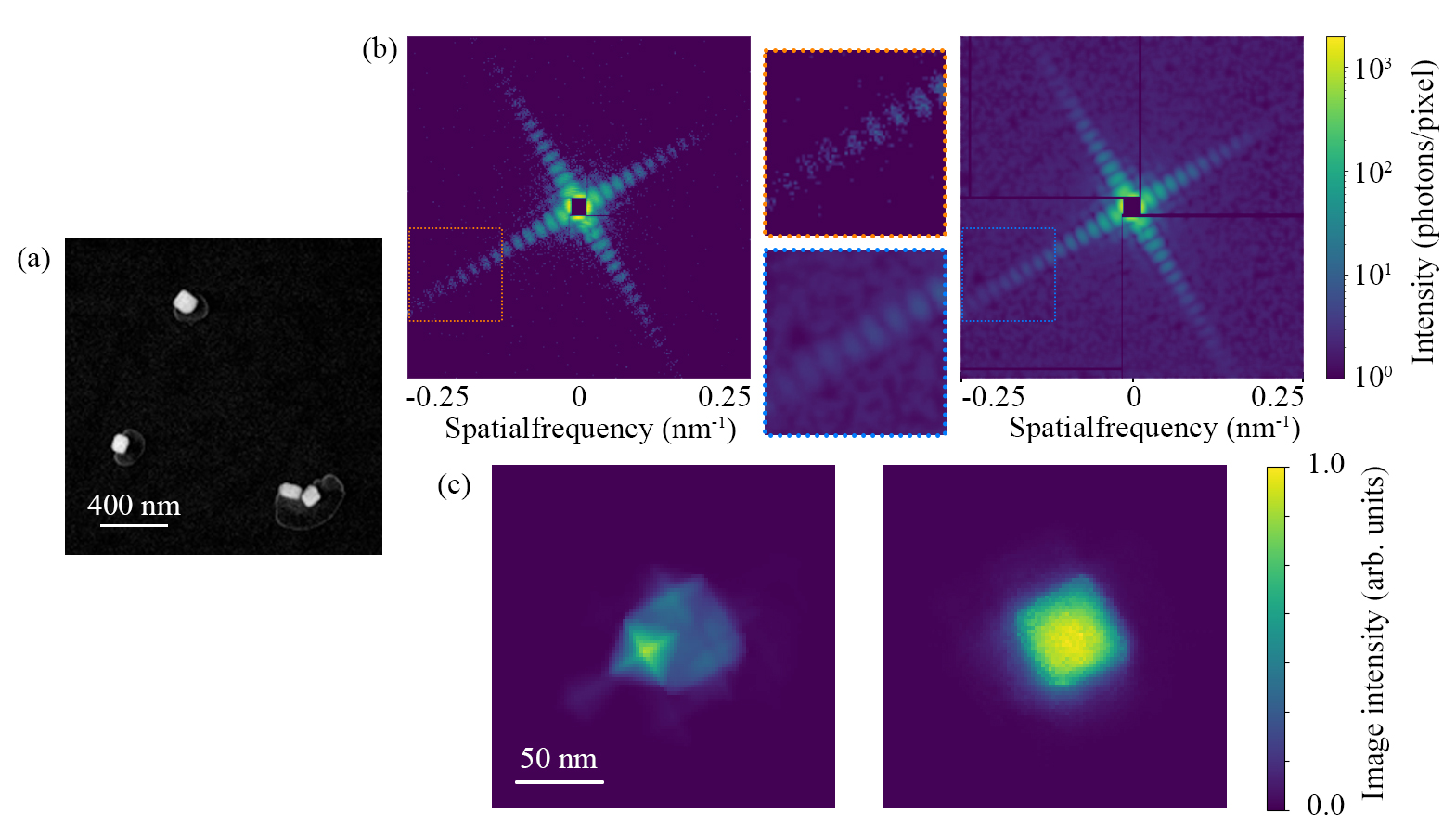}
\caption{\label{fig:ex_noise_reduction}
The results of reconstruction of the experimental diffraction pattern. (a) Scanning Electron Microscope (SEM) image of AgCl nanoparticles. (b) The experimental coherent diffraction pattern from an AgCl nanoparticle with a single XFEL pulse exposure. (c) The denoised experimental data based on the U-Net model with triangle dataset. The color bar represents logarithm intensity. (d) and (e) are the average images of 200 reconstruction calculations from the raw experimental data and the denoised data, respectively.
}
\end{figure}

Figure \ref{fig:ex_noise_reduction} (b) shows the coherent diffraction pattern measured with a single pulse exposure from SACLA. The coherent diffraction pattern was extracted from the center of a 1, 024 $\times$ 1,024 pixel detector and binned into 4 $\times$ 4 pixels to produce 256 $\times$ 256 pixel data. Considering the maximum scattering angle, the edge of the coherent diffraction pattern corresponds to a resolution of 4 nm in real space. The center of the detector is empty to protect the detector from damage caused by the high-intensity direct XFEL beam that is not diffracted by samples. Additionally, the octal sensor of the MPCCD has several seams between  modules resulting in several insensitive regions extending from the center. The incident X-ray pulse energy was about 450 $\mu J$, corresponding to about $10 \times 10^{11 \sim 12}$ photons at 4.0 keV.

Fig.\ref{fig:ex_noise_reduction} (c) shows the denoised image using the pre-trained model based on U-net with triangle dataset. In the enlarged image, we can see that the speckles, which almost disappear in the raw experimental image due to low photon count, are smoothly recovered in the denoised experimental image. It is noted that AgCl nanoparticles basically take a shape similar to a cube. Therefore, the diffraction pattern is also expected to have a clean fringe pattern. We can see that the present denoising analysis makes the fringe pattern clearer.

To retrieve sample images from the experimental coherent diffraction patterns, we used the same algorithms as described in the previous section with the mask processing. We performed 200 runs of image reconstructions from different initial random seeds for each diffraction pattern. Fig.\ref{fig:ex_noise_reduction} (d) and (e) show the averages of all the reconstruction results from the raw and denoised diffraction pattern shown in Fig.\ref{fig:ex_noise_reduction} (b) and (c), respectively. Some artifacts appear in the averaged retrieved image from the raw experimental data, while a solid square is retrieved from the denoised experimental data. The reconstruction from the raw experimental data is also square-shaped to some extent, but the noise can cause it to fall into a false solution, and taking an average of 200 averages does not return to a clean square shape. Conversely, with the denoised experimental data, the solution discrepancies due to noise effects are improved and thus a clean and unique solution was obtained.

\section{Conclusion}
This study addresses the challenge of noise removal in single-shot XFEL-CDI imaging. Using simulation data, we apply a deep-learning approach to image denoising, and show that this leads to improved image reconstruction in real experimental data. Our results highlight the effectiveness of neural network architectures, particularly U-net, which showed remarkable success in recovering complex shaped nanoparticles. Furthermore, by applying transfer learning to the model initially trained with U-net, we offer the capability for efficient model fine-tuning. Our work underscores the potential and utility of incorporating simulation data in deep learning applications for enhanced results.

The techniques present in this research have significant potential in the analysis of X-ray free-electron laser data, particularly for complex molecular or material structure analysis. Improved noise removal, as achieved by our methods, can be crucial when observing phenomena with delicate structures or rapid changes. We believe that the prospective applicability of our methods to X-ray imaging in a wide range of samples and conditions suggests an anticipated expansion in experimental flexibility and scope.

\begin{backmatter}
\bmsection{Funding}
T. I. is supported by International Graduate Program of Innovation for Intelligent World (IIW) of The Univ. of Tokyo. A part of this work was supported by “Advanced Research Infrastructure for Materials and Nanotechnology in Japan(ARIM)” of the Ministry of Education, Culture, Sports, Science and Technology (MEXT), Grant Number JPMXP1223UT1093. This work was supported by JSPS KAKENHI (Grant Number 20H04451, 21K20394, 23H01833, 23KF0019), JST PRESTO (Grant Number JPMJPR1772), Murata Science and Education Foundation, The Precise Measurement Technology Promotion Foundation, and The University of Tokyo Excellent Young Researcher program, Japan.

\bmsection{Acknowledgments}
The XFEL experiments were performed at the BL2 of SACLA (Proposal Numbers 2019B8050, 2020A8044, 2021B8063, and 2022B8027). We thank Y. Matsumoto for the fruitful discussion. We’d like to express our appreciation to Dr. Jordan T. O'Neal for helping to improve the manuscript. In this research work, we used the mdx: Large-Scale Collaborative Platform to Accelerate Data-Driven Science and Society\cite{mdx} and the facilities of the Supercomputer Center, Institute for Solid State Physics, University of Tokyo.

\bmsection{Disclosures}
The authors declare no conflicts of interest.
\bmsection{Data Availability Statement}
All data is uploaded to \url{https://isspns-gitlab.issp.u-tokyo.ac.jp/t-ishikawa/SaclaDenoise}.
\end{backmatter}

%%%%%%%%%%%%%%%%%%%%%%% References %%%%%%%%%%%%%%%%%%%%%%%%%

\bibliography{sample}

\end{document}